\documentclass[conference]{IEEEtran}
\IEEEoverridecommandlockouts
\usepackage{cite}
 \usepackage{tikz}
 \usetikzlibrary{backgrounds}
\usetikzlibrary{fit}        % Add this new one!
\usetikzlibrary{arrows.meta} % Add this new one!

\usepackage{amsmath,amssymb,amsfonts}
\usepackage{algorithmic}
\usepackage{graphicx}
\usepackage{booktabs}
\usepackage{textcomp}
\usepackage{hyperref} 
\usepackage{xcolor}
\usepackage{url}    
\usepackage{lmodern}
\def\BibTeX{{\rm B\kern-.05em{\sc i\kern-.025em b}\kern-.08em
    T\kern-.1667em\lower.7ex\hbox{E}\kern-.125emX}}
\title{Herding CATs: ALARA for Agent Harness Engineering in Portable Composable Multi-Agent Teams}

\author{
	\IEEEauthorblockN{Christopher J. Agostino\IEEEauthorrefmark{1}\IEEEauthorrefmark{2}, Nayan D'Souza\IEEEauthorrefmark{1}\IEEEauthorrefmark{3}}
\IEEEauthorblockA{\IEEEauthorrefmark{1}\textit{NPC Worldwide},
Bloomington, IN, USA,\\
Email: cjp.agostino@gmail.com\\
} 
\IEEEauthorblockA{\IEEEauthorrefmark{2}\textit{Celeria, Inc.},
Bloomington, IN, USA \\
}
\IEEEauthorblockA{\IEEEauthorrefmark{3}\textit{Department of Linguistics, Indiana University},
Bloomington, IN, USA \\
}
}

\begin{document} 

\maketitle

\begin{abstract}
Industry practitioners and academic researchers regularly use multi-agent systems to accelerate their work, but the applications through which users operate these systems do not provide a simple, unified mechanism for scalably managing critical components of the agent harness. This lack of control adversely impacts both the quality of individual human-agent interactions and reduces the capacity for practitioners to coordinate context engineering efforts. The behavioral specifications that define what agents in such systems can do remain fragmented across prose instruction files--for which compliance cannot be guaranteed--or framework-internal configurations, making these specifications difficult to share, version, or collaboratively maintain across teams and projects. Applying the ALARA principle from radiation safety (exposures kept as low as reasonably achievable) to context, we introduce a context-agent-tool (CAT) data layer expressed through interrelated plain-text files, allowing users to directly declare tool access for each agent and to modify the tools themselves that are used by the agents when processing. We demonstrate capability of this CAT data layer to enable real agentic usage by using a command-line shell that loads the team and executes agent runs---\texttt{npcsh}---and evaluating 22 locally-hosted models from 0.6B to 35B parameters across 115 practical tasks spanning file operations, web search, multi-step scripting, tool chaining, and multi-agent delegation. We characterize which model families succeed in certain task categories and where they break down across $\sim$2500 total executions. 
\end{abstract}
\begin{IEEEkeywords}
human-agent interaction, multi-agent systems, composable automation, agent evaluation, co-creative systems
\end{IEEEkeywords}

\section{Introduction}

Multi-agent systems built on large language models have become commonplace tools for practitioners across software development, research, creative work, and more. As these systems have grown more sophisticated, the infrastructure surrounding them (the agent harness that governs tool access, behavioral boundaries, and inter-agent coordination) has become a distinct engineering concern~\cite{anthropic2024agents}. Users who want to specify what their agents can do, what tools they have access to, and how they should coordinate currently express these specifications through a combination of prose instruction files (Cursor's \texttt{.cursorrules}, Claude Code's \texttt{CLAUDE.md}, Aider's convention files~\cite{aider}, the emerging \texttt{AGENTS.md} standard), programmatic framework code (LangGraph~\cite{langchain}, AutoGen~\cite{autogen}, CrewAI~\cite{crewai}), and tool registration protocols like MCP servers. Each of these mechanisms addresses an important piece of the agent harness, but they do not compose with each other. Prose instructions rely on interpretive compliance that cannot be guaranteed in principle, given that meaning-production in language models is contextual in the formal sense, generated in the act of interpretation rather than retrieved from pre-existing associations~\cite{agostino2025qnlp,agostino2026bell,trouillas2024,thomas2026}. On the more programmatic front, the most popular frameworks for agents tend to embed behavioral specifications within core application logic (most typically the core system prompt) which can be difficult to inspect or modify without deep expertise. MCP servers, while they enable users the ability to quickly add useful tools to their agents' arsenal, scope tool access to the server at large rather than allowing users to clearly specify which individual agents in their multi-agent systems have access to specific tools. As a result, the aspects of the harness most consequential for the quality of human-agent interaction are fragmented across mechanisms that either cannot be validated programmatically or are too technically difficult for subject-matter experts to collaboratively maintain without substantial investment or dedicated engineering effort.

The cost of this fragmentation of critical agent infrastructure has accumulated from several directions. Language model performance on retrieval tasks degrades significantly when relevant information appears in the middle of the context window, with the lowest accuracy at precisely the positions where accumulated conversation history pushes the behavioral instructions that an agent most needs to follow~\cite{liu2023lost}. An analysis of 1{,}600 execution traces across seven multi-agent frameworks found that specification and system design failures account for 44\% of all breakdowns, making underspecification, rather than model capability or network failures, the plurality failure mode~\cite{cemri2025mast,chan2023harms}. From the security side, work on privilege control for LLM agents has identified over-privileged tool access as the structural enabler of prompt injection attacks: agents given access to tools they do not need for a given role create attack surface that injected content can exploit~\cite{shi2025progent}. In a system that combines untrusted data with privileged tools, least privilege enforcement is necessary but inherently unreliable when implemented through prose instructions~\cite{pfi2025}. The plan-then-execute pattern resists injection because the agent commits to a tool plan before processing untrusted content, leaving no interpretive surface for injected frames~\cite{tramer2025patterns}. These empirical findings demonstrate that specifications-as-prose for multi-agent systems are fundamentally limited in preventing capability degradation or security vulnerabilities.

The principle that the exposure of a system to risk should be kept as low as reasonably achievable, codified by the International Commission on Radiological Protection in 1977~\cite{icrp1977} and formalized by the US Nuclear Regulatory Commission as a continuing optimization toward the minimum that the purpose of the activity actually demands~\cite{nrc_alara}, predates the modern internet but maps directly onto the problem of agent tool access, where its structural analog in computer security (least privilege, the restriction of access to the minimum necessary to accomplish assigned tasks~\cite{nist_polp}) has already been identified as the necessary response to the over-privileged tool catalogs that enable prompt injection~\cite{shi2025progent,pfi2025,tramer2025patterns}. The constraint in both formulations holds regardless of what the agent interprets or misinterprets from its context, because tools not present in an agent's schema cannot be invoked. Experiments on tool catalog size confirm the practical stakes of this requirement as tool invocation accuracy falls from $\sim$95\% to $\sim$25\% as catalog size grows from one to eight~\cite{shen2023taskbench}, with the degradation concentrated in the space between what a model is given access to and what it should actually be using.

The accumulation of transient, unauditable configuration artifacts across agentic systems~\cite{xu2025context} resembles the problem that Infrastructure as Code solved for deployment through the introduction of declarative specification files that a system could parse and enforce rather than prose that a human would read and enact. In this work, our principal contribution is a declarative context-agent-tool (CAT) data layer that expresses the agent harness through interrelated files scoping each agent's tool access and context to the minimum its role requires. To demonstrate its capabilities in enabling agent harnesses, we use the open-source tool command-line shell \texttt{npcsh} to evaluate it on a suite of benchmark tasks designed for smaller language models and specifically to test agentic capabilities.

The remainder of the paper describes the design of the framework (Section~\ref{sec:design}) and presents a 115-task benchmark evaluated across 22 locally-hosted models from 0.6B to 35B parameters, with trace-level analysis of 2{,}530 task executions characterizing when and how agent reliability breaks down across model families, task categories, and retry strategies (Section~\ref{sec:benchmark}).

\section{Design and Methods}
\label{sec:design}

Scoping each agent's tool access and context to the minimum its role requires demands a specification format that a system can parse and enforce structurally rather than one that relies on interpretive compliance. We describe below the declarative data layer through which we represent the agent harness (Section~\ref{sec:cat}) and the file types that compose it (Sections~\ref{sec:jinxes}--\ref{sec:skills}).

\subsection{The CAT Data Layer}
\label{sec:cat}

\begin{figure}[h]
\centering
\resizebox{\columnwidth}{!}{%
\begin{tikzpicture}[
  npc/.style={rectangle, draw, fill=blue!10, minimum width=1.3cm, minimum height=0.4cm, font=\scriptsize},
  ctx/.style={rectangle, draw, fill=orange!15, rounded corners, minimum width=1.3cm, minimum height=0.4cm, font=\scriptsize},
  jinx/.style={rectangle, draw, fill=green!10, minimum width=1.1cm, minimum height=0.4cm, font=\scriptsize},
  team/.style={rectangle, draw, dashed, inner sep=4pt},
  uses/.style={-{Stealth[length=3pt]}, semithick, green!50!black},
  assign/.style={-{Stealth[length=3pt]}, thin, blue!50, dashed}
]
% ---- Teams (left) ----
\node[ctx] (ctx1) at (0, 0) {context};
\node[npc] (orch) at (0, -0.65) {orchestrator};
\node[npc] (a1) at (0, -1.3) {agent\_a};
\node[npc] (a2) at (0, -1.95) {agent\_b};
\node[ctx] (ctx2) at (0, -2.9) {context};
\node[npc] (suborch) at (0, -3.55) {sub\_orch};
\node[npc] (spec) at (0, -4.2) {specialist};
\begin{scope}[on background layer]
  \node[team, fit=(ctx2)(suborch)(spec), label={[font=\tiny]above left:sub-team/}] (st) {};
  \node[team, fit=(ctx1)(orch)(a1)(a2)(st), label={[font=\tiny]above left:team/}] (tt) {};
\end{scope}
% ---- Jinx DAG (right) ----
% ---- Jinx DAG: root at top, composed builds downward ----
% Level 0: root
\node[jinx] (python) at (5.3, 0.3) {python};
% Level 1: inherit from python
\node[jinx] (chat) at (3.5, -0.7) {chat};
\node[jinx] (sh) at (5.0, -0.7) {sh};
\node[jinx] (websearch) at (6.5, -0.7) {web\_search};
\node[jinx] (screenshot) at (8.0, -0.7) {screenshot};
% Level 1 edges
\draw[uses] (python) -- (chat);
\draw[uses] (python) -- (sh);
\draw[uses] (python) -- (websearch);
\draw[uses] (python) -- (screenshot);
% Level 2: composed Jinxes built from level-1 steps
\node[jinx] (react) at (4.2, -1.8) {react};
\node[jinx] (delegate) at (5.6, -1.8) {delegate};
\node[jinx] (compuse) at (7.3, -1.8) {computer\_use};
% Level 2 edges: shared parents show DAG reuse
\draw[uses] (chat) -- (react);
\draw[uses] (python) to[out=-70,in=120] (react);
\draw[uses] (chat) -- (delegate);
\draw[uses] (sh) -- (delegate);
\draw[uses] (chat) -- (compuse);
\draw[uses] (screenshot) -- (compuse);
\draw[uses] (sh) -- (compuse);
% ---- Assignment ----
\draw[assign] (react.west) to[out=180,in=0] (orch.east);
\draw[assign] (websearch.west) to[out=180,in=20] (a1.east);
\draw[assign] (compuse.west) to[out=190,in=0] (a2.east);
\draw[assign] (delegate.west) to[out=190,in=0] (suborch.east);
\draw[assign] (react.west) to[out=195,in=0] (spec.east);
\end{tikzpicture}
}%
\caption{The CAT data layer. \textit{Left:} context files (orange) scope teams of NPCs (blue). \textit{Right:} Jinxes (green) compose as a DAG rooted at \texttt{python}. Level-1 Jinxes (\texttt{chat}, \texttt{sh}, \texttt{web\_search}, \texttt{screenshot}) each use \texttt{python} as their engine. Level-2 Jinxes combine level-1 steps: \texttt{react} chains \texttt{chat}~+~\texttt{python}, \texttt{computer\_use} chains \texttt{chat}~+~\texttt{screenshot}~+~\texttt{sh}, and \texttt{delegate} uses \texttt{chat}~+~\texttt{sh}. Shared parents (e.g.\ \texttt{chat} used by all three composites) make the graph a DAG rather than a tree. Dashed arrows show tool-catalog assignment to NPCs.}
\label{fig:cat}
\end{figure}
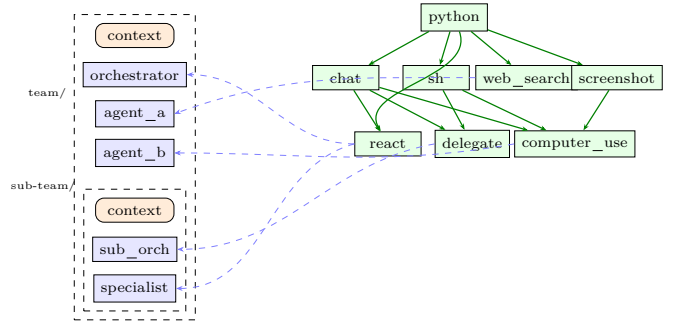

We represent the agent harness through three interrelated file types: context files that scope shared resources and designate orchestrators, NPC files that define individual agents with their model configurations and tool permissions, and Jinxes (Jinja Execution templates) that specify the tools themselves as executable YAML definitions. The relationships among these file types are shown in Figure~\ref{fig:cat}. Context files sit at the team level and determine which agent serves as orchestrator and what sub-teams exist within the hierarchy. Each NPC file references a subset of Jinxes from the available catalog, and because Jinxes can invoke other Jinxes as execution steps, the tool layer forms a directed acyclic graph of composable capabilities. Because the entire specification consists of files on disk rather than framework-internal state, the same directory transfers without modification across every interface the system provides, including an interactive shell (\texttt{npcsh}), an API server (\texttt{npc serve}), a desktop application (\texttt{npcts}), a browser-based IDE (\texttt{incognide}), and direct Python import.

\subsection{Jinxes}
\label{sec:jinxes}

A Jinx is a YAML file specifying a name, a natural-language description, typed inputs, and a sequence of execution steps, where each step names an engine (\texttt{python}, \texttt{bash}, \texttt{llm}, or another Jinx) and contains Jinja-templated code rendered with input values at execution time. We derive tool-calling schemas directly from the Jinx's inputs and description, so the file serves as both the definition and the executable artifact, eliminating the schema-definition layer where drift between specification and behavior accumulates in programmatic frameworks.

Because tool-use capability varies substantially across model families and cannot be assumed from parameter count or benchmark performance on other tasks~\cite{schick2023toolformer}, we use the Jinx to provide deterministic scaffolding (execution order, data flow between steps, error handling) that bounds the interpretive surface to individual steps where model-driven interpretation adds value. Where the ReAct pattern treats reasoning traces and tool actions as a unified loop~\cite{yao2023react}, we decompose this loop into a deterministic skeleton whose interpretive demands are scoped to individual steps. When a step names another Jinx as its engine, that Jinx expands in place with arguments substituted and results threading through shared context, so that single-agent workflows compose into directed acyclic execution graphs from individually simple files (Figure~\ref{fig:cat}, green nodes). In practice, a research pipeline chains hypothesis generation, sub-agent search, iterative writing, and review-revision cycles as a single Jinx whose steps invoke other Jinxes three levels deep, and a desktop automation workflow chains screenshot capture, vision analysis, and action execution in a deterministic loop. In both cases the user who wants to understand what the system does reads the top-level file, and the user who wants to change it edits one step. Because Jinxes execute through prompt-based flows rather than requiring native function-calling support, they work with any model regardless of whether it implements structured tool calling, which enables agent workflows on small locally-hosted models that would otherwise be excluded from tool-using architectures entirely. The same Jinx that a user invokes as a slash command serves as the tool-call target when an agent selects it during reasoning, presenting an interactive interface when a human invokes it directly and executing the same code with the same parameters when an agent calls it autonomously~\cite{horvitz1999}. A correction made to a Jinx propagates to every agent that uses it, providing the kind of unified human-agent tool definition that declarative agent specification has been identified as requiring~\cite{zeng2025adl}. We organize Jinxes in a categorized directory hierarchy and reference them by name through template resolution rather than by path, so that reorganizing the capability taxonomy does not break agents when compiled at runtime.

\subsection{NPCs}
\label{sec:npcs}

An NPC file defines an agent through a name, a natural-language directive, a model and provider specification, and a Jinx list that simultaneously constitutes the tool catalog and the permission set. We use the Jinx list to scope each agent to the minimum set of tools its role requires, an intervention motivated by the finding that tool selection accuracy undergoes a structural phase transition as catalog size grows~\cite{shen2023taskbench,li2026single} and that agents operating with the smallest set of high-signal tokens outperform those given overlapping tools~\cite{anthropic2024agents}.

The Jinx list enforces its constraint structurally rather than interpretively. A prose instruction file can request that an agent limit itself to certain tools, but the model is free to ignore or misinterpret the instruction, and attention dilution makes this increasingly likely over long contexts~\cite{liu2023lost}; tools not on the Jinx list do not exist in the agent's schema, and no amount of prompt injection or attention drift can invoke what was never provided. The security literature arrives at this conclusion through analysis of attack surfaces~\cite{shi2025progent,tramer2025patterns,pfi2025}, and we implement it through the definition format rather than through a separate enforcement layer.

The same mechanism operationalizes autonomy as a continuous design parameter. An NPC with a narrow Jinx list and no delegation tool operates at low autonomy, one with the delegation Jinx and a broad catalog operates at higher autonomy, and the user repositions any agent on this spectrum by editing a YAML list. That high automation and high user control are independently achievable rather than opposed~\cite{shneiderman2022hcai} is what the NPC file makes concrete, operationalizing the automation taxonomy~\cite{parasuraman2000} and the autonomy levels that require explicit tradeoffs at each stage~\cite{feng2025autonomy} through a mechanism simple enough that users actually edit it, addressing the gap that existing co-creative systems leave open~\cite{zhang2025agency}.

\subsection{Teams}
\label{sec:teams}

A team is a directory containing NPC files, a Jinxes subdirectory, and a context file designating the orchestrator and configuring shared resources. Role specialization and focused context produce gains across multi-agent systems~\cite{hong2024metagpt,chen2024agentverse,liu2024dylan,qian2024chatdev}, and separating planning, calling, and summarizing into focused roles substantially improves performance in the sub-24B range~\cite{shen2024small}; in this framework, decomposition is reorganization, because moving NPC files into subdirectories, each with its own context file and orchestrator, creates sub-teams whose internal complexity is invisible to the top-level router. When the top-level orchestrator routes to a sub-team it sees only the sub-team's description rather than the full persona descriptions and tool catalogs of every agent within it, keeping the routing decision tractable for small models whose context budgets are constrained and preventing context from the wrong scope from entering the routing decision. Loss of conversation history is a top failure mode in existing multi-agent frameworks~\cite{cemri2025mast}, and the sub-team boundary addresses it by construction. We implement delegation between NPCs as a Jinx rather than a framework primitive, so that the user modifies completion criteria, feedback mechanisms, and iteration limits by editing a file, and the same delegation Jinx serves both human-invoked and agent-invoked paths.

\subsection{Skills}
\label{sec:skills}

Skills are Jinxes that deliver instructional content rather than execute code, providing agents with methodology and domain knowledge through the same tool mechanism used for any other capability. Because they are Jinxes, skills appear in agent tool catalogs alongside executable tools and are assigned through the same Jinx list. We implement section-level retrieval within skills so that an agent requests only the portion of a methodology document it needs rather than loading the full file into context, managing the token budget under which contextual interpretation must operate.

\subsection{Benchmark}
\label{sec:benchmark}

Existing agentic benchmarks (SWE-bench~\cite{swebench}, HumanEval~\cite{humaneval}, Terminal-Bench~\cite{terminalbench}, AgentBench~\cite{agentbench}) report scores that conflate model reasoning with framework orchestration~\cite{kapoor2024agents}, are highly sensitive to unreported implementation details~\cite{biderman2024lessons}, and in the case of SWE-bench reflect data quality problems in a substantial fraction of passes~\cite{aleithan2024swebenchplus}. Enterprise applications introduce requirements (high reliability, auditable traces, complex interaction patterns) that existing benchmarks rarely address~\cite{agenteval2025kdd}, and benchmarks calibrated for frontier models produce uniformly low scores against 1B--35B models, obscuring whether the framework's tool orchestration and error recovery are functioning at all.

We designed a benchmark of 115 tasks spanning 13 capability categories, each specifying a natural-language instruction and a verification command that checks filesystem state. Tasks range from single-tool invocations through multi-step sequencing to delegation chains and tool composition, extending beyond code-focused tasks to include web search, multi-agent delegation, and tool chaining. The design is framework-agnostic: the same instructions and verification commands can evaluate any agent system that accepts natural-language input, enabling cross-framework comparison holding the model constant.

\section{Results}
\label{sec:results}

\begin{figure}[h]
\centering
\includegraphics[width=\columnwidth]{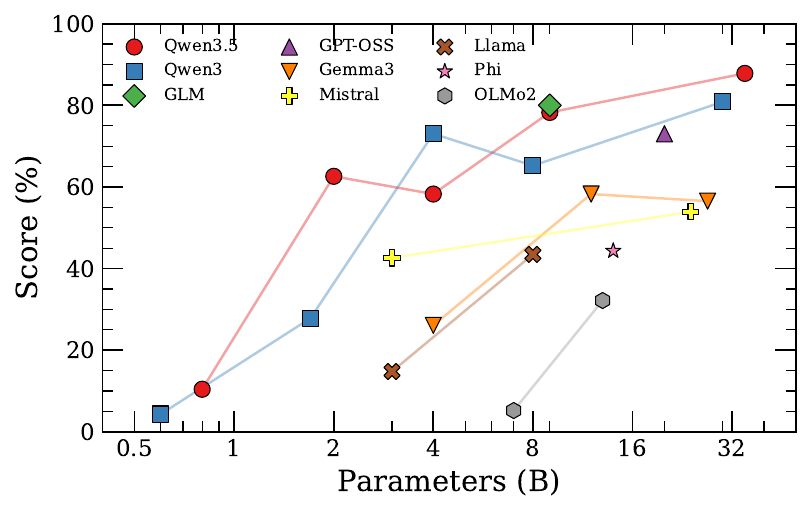}
\caption{Benchmark score versus parameter count across model families.}
\label{fig:scaling}
\end{figure}

We evaluated 22 locally-hosted models from 9 model families, spanning 0.6B to 35B parameters, all running via Ollama on consumer hardware with a 360-second timeout per task and up to 5 retry attempts with error feedback. Table~\ref{tab:results} presents overall scores and Figure~\ref{fig:scaling} shows score as a function of parameter count. We extract per-task trace data from the 2{,}530 task executions: attempt counts, tool call counts, and task durations split by success and failure. Cloud-hosted frontier models (GPT-4o, Claude Sonnet, Gemini Pro) score above 90\% on these same tasks, confirming that the benchmark measures model capability rather than limitations of the declarative scaffolding.

\begin{table}[h]
\caption{Benchmark results across model families.}
\label{tab:results}
\begin{tabular}{llrc}
\toprule
Family & Model & Params & Score \\
\midrule
Qwen3.5 & 0.8b & 0.8B & 12/115 (10\%) \\
        & 2b   & 2B   & 72/115 (63\%) \\
        & 4b   & 4B   & 67/115 (58\%) \\
        & 9b   & 9B   & 90/115 (78\%) \\
        & 35b  & 35B  & 101/115 (88\%) \\
\midrule
Qwen3   & 0.6b & 0.6B & 5/115 (4\%) \\
        & 1.7b & 1.7B & 32/115 (28\%) \\
        & 4b   & 4B   & 84/115 (73\%) \\
        & 8b   & 8B   & 75/115 (65\%) \\
        & 30b  & 30B  & 93/115 (81\%) \\
\midrule
GLM     & 4.7-flash & 9B & 92/115 (80\%) \\
GPT-OSS & 20b  & 20B  & 84/115 (73\%) \\
\midrule
Gemma3  & 4b   & 4B   & 30/115 (26\%) \\
        & 12b  & 12B  & 67/115 (58\%) \\
        & 27b  & 27B  & 65/115 (57\%) \\
\midrule
Mistral & small3.2 & 24B & 62/115 (54\%) \\
        & ministral-3 & 3B & 49/115 (43\%) \\
\midrule
Llama   & 3.2:3b & 3B & 17/115 (15\%) \\
        & 3.1:8b & 8B & 50/115 (43\%) \\
\midrule
Phi     & phi4 & 14B  & 51/115 (44\%) \\
\midrule
OLMo2   & 7b   & 7B   & 6/115 (5\%) \\
        & 13b  & 13B  & 37/115 (32\%) \\
\bottomrule
\end{tabular}
\end{table}

Within-family scaling is monotonic, but between-family differences dominate (Figure~\ref{fig:scaling}). The between-family variance exceeds the within-family variance at every parameter count where multiple families overlap, indicating that the quality of tool-use training varies enough across families to overwhelm order-of-magnitude differences in parameter count. Tool use is a trained capability~\cite{schick2023toolformer}, and these results show that models trained for it achieve scores at 4B that models not trained for it fail to reach at 27B. Models below 3B can parse the agentic prompt and occasionally select the correct tool but cannot reliably form arguments or sequence multi-step operations.

\begin{figure}[h]
\centering
\includegraphics[width=\columnwidth]{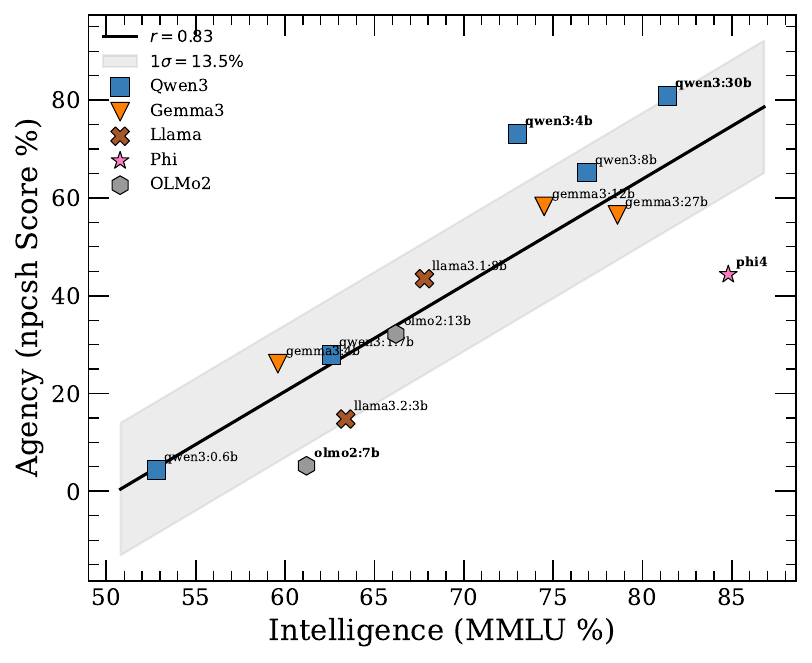}
\caption{Agency (npcsh score) versus intelligence (MMLU). The regression line ($r \approx 0.8$) with $1\sigma$ band; outliers beyond $1\sigma$ are labeled.}
\label{fig:mmlu}
\end{figure}

Cross-referencing benchmark scores with MMLU reveals a linear relationship ($r \approx 0.8$, $p < 10^{-3}$) between general capability and agentic performance (Figure~\ref{fig:mmlu}). The outliers from this trend identify models whose training transfers disproportionately well or poorly to tool-use scenarios, consistent with the BFCL finding that single-turn accuracy and multi-turn reliability diverge in ways general benchmarks do not predict~\cite{patil2025bfcl}.

\begin{figure}[h]
\centering
\includegraphics[width=\columnwidth]{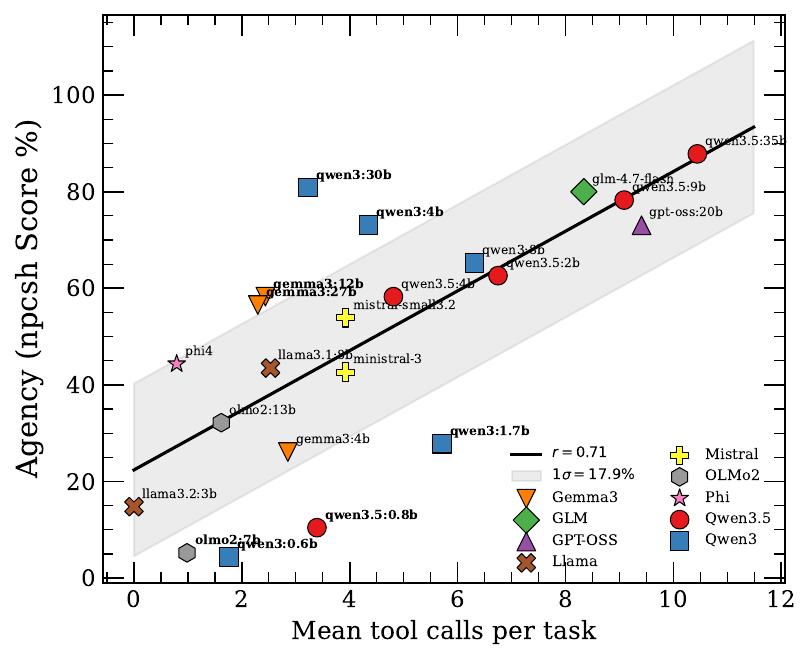}
\caption{Mean tool calls per task versus agency score ($r \approx 0.7$).}
\label{fig:toolcalls}
\end{figure}

The mean number of tool calls per task is a stronger predictor of benchmark score ($r \approx 0.7$, $p < 10^{-3}$) than either mean task duration ($r \approx 0.3$) or mean attempt count ($r \approx 0.5$), shown in Figure~\ref{fig:toolcalls}. Higher-scoring models make more tool calls rather than spending more time or retrying more often. The relationship between tool use on successful versus failed tasks reveals two behavioral clusters: models that use substantially more tools when failing (persisting without convergence) and models that use more tools when succeeding (genuine multi-step problem solving with fast failure exits). Pass rates below 50\% for leading models on realistic multi-turn tasks~\cite{yao2024taubench} indicate that reliability under sustained tool engagement is the operative bottleneck.

\begin{figure}[h]
\centering
\includegraphics[width=\columnwidth]{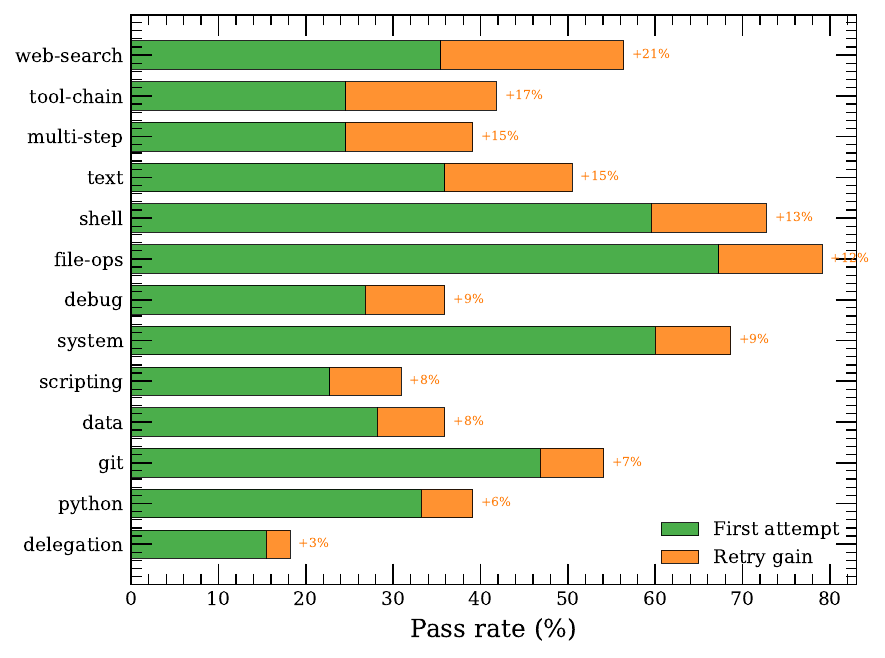}
\caption{First-attempt pass rate (green) and additional gain from retries (orange) by task category.}
\label{fig:retry}
\end{figure}

Across all models, $\sim$80\% of successful task completions occur on the first attempt. The correlation between first-attempt success rate and retry gain is positive ($r \approx 0.5$, $p < 0.05$), suggesting that retries help only when the model can learn from error feedback. Retry value varies by an order of magnitude across task categories (Figure~\ref{fig:retry}): web search gains $\sim$20 percentage points from retries, tool-chain $\sim$15, multi-step and text $\sim$15 each, while delegation gains $\sim$3. For categories where the marginal gain per retry is low, the accumulated context from prior failures may actively degrade performance through loss of conversation history, failure mode 1.4 in the MAST taxonomy~\cite{cemri2025mast}.

\begin{figure*}[t]
\centering
\includegraphics[width=\textwidth]{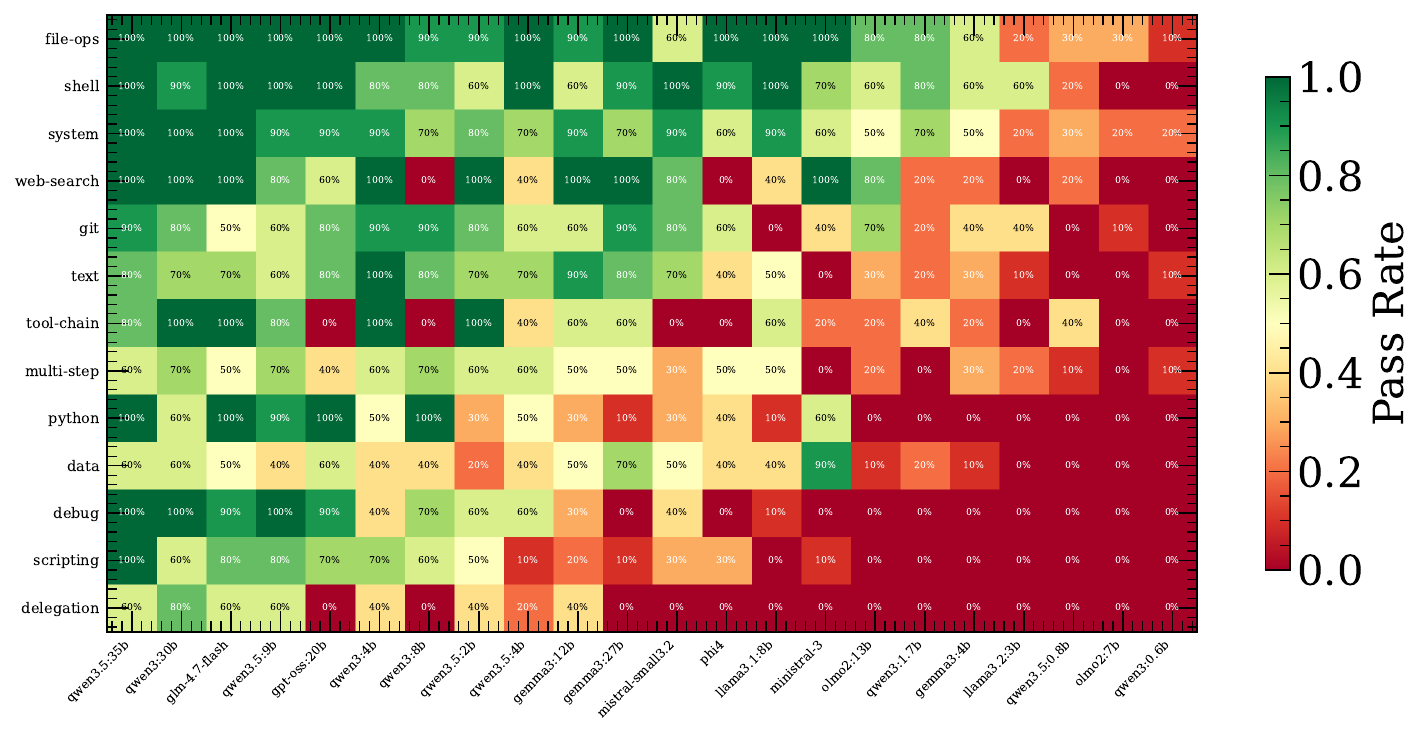}
\caption{Pass rate by model and task category. Models ordered by overall score (left to right); categories by mean difficulty (top to bottom).}
\label{fig:famcat}
\end{figure*}

The 13 task categories span from file operations ($\sim$80\% mean pass rate) and shell tasks ($\sim$75\%) through web search ($\sim$55\%) to scripting ($\sim$30\%) and delegation ($\sim$20\%). The difficulty ordering holds across model families but the magnitude of degradation varies (Figure~\ref{fig:famcat}). Delegation is the hardest category for every model, and categories requiring multi-step reasoning are difficult across all families while categories requiring only correct tool invocation are easy.

\section{Discussion}

The between-family variance in agentic capability, which exceeds the within-family variance at every overlapping parameter count, confirms that tool-use reliability is a distinct trained capability that general benchmarks underpredict~\cite{patil2023gorilla,patil2025bfcl}. The same pattern has been found in the sub-24B range \cite{shen2024small}, where separating planning, calling, and summarizing into focused roles produced substantial gains. Our delegation results, which show the highest failure rate and the lowest retry gain of any category, are consistent with the 44\% specification-and-design failure rate in the MAST taxonomy~\cite{cemri2025mast}, suggesting that the failures we observe in delegation are primarily specification failures rather than capability failures.

The category-level variation in retry value has practical implications for retry policy. For categories where accumulated context from prior failures actively degrades performance (failure mode 1.4 in~\cite{cemri2025mast}), the rational policy is to restart with a clean context rather than retry. The positive correlation between first-attempt success rate and retry gain ($r \approx 0.5$) is opposite to what a ceiling-effect model would predict, and suggests that retries help only when the model has sufficient baseline capability to learn from error feedback. The $\tau$-bench results support this interpretation: pass rates below 50\% under sustained tool engagement indicate a reliability bottleneck distinct from single-call accuracy~\cite{yao2024taubench}.

The Jinx-list scoping approach extends the declarative specification work of DSPy~\cite{khattab2024dspy} and ADL~\cite{zeng2025adl} through a shell interface and filesystem organization. That the context engineering~\cite{xu2025context} and Unix philosophy~\cite{piskala2026unix} literatures arrive at the same architectural argument independently supports the IaC analogy. Independent security analyses~\cite{shi2025progent,pfi2025,tramer2025patterns} each arrive at tool-level constraint enforcement as a necessary mechanism; the connection to the ICRP's ALARA formulation~\cite{icrp1977} and NIST's least privilege~\cite{nist_polp} is structural rather than metaphorical, as both describe the same optimization of minimizing exposure given the understanding that the exposure over time will lead to some kind of degradation. In the case of LLMs and AI agents, the degradation is severe and well documented at large contexts.

The co-creative design of the NPC file, in which the Jinx list is both the tool catalog and the permission set, addresses gaps identified in concrete mechanisms for shared agency \cite{zhang2025agency}, following the mixed-initiative framework of \cite{horvitz1999} and the finding of \cite{shneiderman2022hcai} that high automation and high user control are jointly achievable. Guidelines call for transparency and correctability \cite{amershi2019}; the Jinx list provides both through a file the user can read and edit. For organizational deployment, the filesystem-organized configuration addresses lifecycle management requirements identified \cite{lopes2026clinical,kreuzberger2023mlops}.

The degradation we observe as tool catalog size increases is predicted by the contextuality established by semantic Bell inequality violations in LLMs~\cite{agostino2025qnlp,agostino2026bell,trouillas2024,thomas2026}, which show that meaning-production in these systems is generated in the act of interpretation rather than retrieved from pre-existing associations. These results provide an information-theoretic basis for why prose-based behavioral constraints cannot be guaranteed to produce the intended behavior regardless of scale, and why structural enforcement at the schema level is not only necessary but beneficial for outcomes.

\section{Conclusions}

In this work, we introduced a declarative context-agent-tool (CAT) data layer for scoping agent tool access and context, and evaluated it across 22 locally-hosted models and 2{,}530 task executions. We summarize the main results.

\begin{enumerate}
\item The CAT data layer, consisting of context files, NPC definitions, and Jinxes, enforces tool access constraints structurally rather than interpretively. Tools not present in an agent's Jinx list do not exist in its schema and cannot be invoked regardless of prompt content.

\item Between-family differences in agentic capability dominate within-family scaling, confirming that tool-use reliability is a distinct trained capability. Models trained for tool use achieve scores at 4B parameters that models not trained for it fail to reach at 27B.

\item MMLU and agentic performance correlate ($r \approx 0.8$), but outliers identify models whose training transfers disproportionately well or poorly to tool-use scenarios.

\item Tool call volume is the strongest predictor of agentic performance ($r \approx 0.7$), stronger than task duration ($r \approx 0.3$) or attempt count ($r \approx 0.5$).

\item $\sim$80\% of successful completions occur on the first attempt. Retry value varies by an order of magnitude across task categories, from $\sim$20 percentage points for web search to $\sim$3 for delegation, arguing for category-aware resource allocation.

\item Delegation is the hardest category for every model. The low retry gain for delegation suggests that accumulated context from prior failures degrades rather than aids convergence.

\item The CAT data layer can be fundamentally expressed or compiled in any suitable programming language, and we provide core libraries in rust\footnote{https://github.com/NPC-Worldwide/npcrs} and in python\footnote{https://github.com/NPC-Worldwide/npcpy}, with a typescript \footnote{https://github.com/NPC-Worldwide/npcts} version in progress. The command-line shell that provides runners for both the rust and python runtimes---as well as the benchmarking dataset and scripts---is npcsh\footnote{https://github.com/NPC-Worldwide/}. We also provide a full IDE (incognide \footnote{https://github.com/NPC-Worldwide/incognide}) that allows agents to conduct computer use through the CAT data layer is available, with future work planned for benchmarking it across a suite a local of models.
\end{enumerate}

\bibliographystyle{IEEEtran}
\bibliography{refs}

\end{document}